# Surface passivation for narrowing optical linewidth of silicon T centers in nanophotonic devices

Fariba Islam[1,2,3], Chang-Min Lee[1,2,*], Kyu-Young Kim[1,2,4], Sorah Fischer[1,2,3], Purbita Purkayastha[1,2,5], Amirehsan Alizadehherfati[1,2,3], and Edo Waks[1,2,3,5,*]

[1] Institute for Research in Electronics and Applied Physics, University of Maryland, College Park, Maryland 20742, USA
[2] Joint Quantum Institute, University of Maryland, College Park, Maryland 20742, USA
[3] Department of Electrical and Computer Engineering, University of Maryland, College Park, Maryland 20740, USA
[4] Department of Physics, Ulsan National Institute of Science and Technology, Ulsan 44919, Republic of Korea
[5] Department of Physics, University of Maryland, College Park, Maryland 20742, USA

**Abstract:**
Silicon T centers are promising solid–state spin–photon interfaces for telecom–compatible, scalable quantum information technologies. A major challenge for T centers in nanophotonics is spectral diffusion, where local electric–field environment fluctuations from nearby charge states broaden the optical transition, reducing photon indistinguishability. Strategies that directly suppress spectral diffusion are therefore critical for improving T-center-based quantum devices. Here, we use atomic–layer–deposited $Al_2O_3$ to passivate the silicon surface and demonstrate a systematic narrowing of T center optical linewidths. Across our measurements, $Al_2O_3$ passivation reduces the T center emission linewidth by up to 57%. Complementary above-bandgap illumination and spectral hole burning measurements show that the remaining linewidth contains a significant spectral-diffusion component from adjacent charge traps, while placing an upper bound of 75 MHz on the homogeneous linewidth. This work provides a CMOS–compatible path toward generating indistinguishable photons from silicon T centers for scalable quantum photonic applications.

Silicon T centers have emerged as promising solid–state spin–photon interfaces for scalable quantum information technologies, combining an optically addressable ground-state spin with telecom-band optical emission in a host material compatible with mature, foundry-scale silicon fabrication[1,2]. This emitter provides optically accessible electron spin ground states coupled to intrinsic nuclear spins, with both electron and nuclear spins exhibiting long spin coherence times [1,3]. Development of this color center has advanced rapidly, with demonstrations of nanophotonic integration[3–10], coherent electron and nuclear spin control[3,10], and detailed investigations of its optical, electronic, and spin properties[1–12]. Beyond continued improvement of spin properties, generating indistinguishable photons is a prerequisite for a coherent, scalable spin–photon interface [7].

Optical linewidth broadening of the T centers poses a major challenge for generating indistinguishable photons. The Fourier-transform-limited linewidth of the silicon T center optical transition is approximately 170 kHz [1] and homogeneous broadening has been reported to be 67 MHz in nanofabricated SOI [4]. In practice, however, T centers integrated in nanophotonic devices exhibit far broader optical linewidths of 1–5 GHz [5–7], indicating that spectral diffusion, rather than homogeneous broadening, is the dominant mechanism. Spectral diffusion occurs when slow fluctuations in the local electric-field environment shift the optical transition frequency over time. Such fluctuations can arise from surface charge states including dangling bonds and dopant acceptors, or charge traps in the silicon crystal[2,6,7,13–15] and are particularly severe in T centers integrated with nanophotonics where the emitters are located close to surfaces. Conditional photoluminescence excitation has recently been demonstrated as a means of recovering narrow optical linewidths through post-selection,[6,7] but it does not suppress the underlying spectral diffusion. Therefore, strategies that directly mitigate the physical sources of spectral diffusion are essential.

Surface passivation has improved the optical performance of several solid-state emitter platforms, including diamond color centers[16,17] and semiconductor quantum dots[18], by reducing surface-induced charge noise and enhancing linewidth, spectral stability, or quantum efficiency. However, its impact on silicon T centers remains unexplored. This motivates the use of $Al_2O_3$, a dielectric widely employed in silicon microelectronics[19] and photovoltaics[20–22] to passivate dangling bonds and suppress surface recombination, as a promising strategy to reduce spectral diffusion and narrow the optical linewidth of T centers.

Here, we demonstrate that $Al_2O_3$ surface passivation reduces the optical linewidth of silicon T centers integrated in nanophotonic devices. Using atomic-layer-deposited $Al_2O_3$ as a passivation layer, we observe linewidth narrowing in both same-emitter comparisons and statistical measurements across multiple emitters. In the same-emitter measurements, $Al_2O_3$ passivation reduces the optical linewidth by 32% on average, with a maximum reduction of 57%. We identify 11 nm as an optimal passivation thickness from the statistical measurements. Above-bandgap illumination and spectral hole burning provide complementary probes of the residual broadening after surface passivation, indicating contributions from both charge trap states and homogeneous broadening. These results establish $Al_2O_3$ as an effective, foundry-compatible passivation layer for mitigating spectral diffusion in silicon T-center nanophotonic devices.

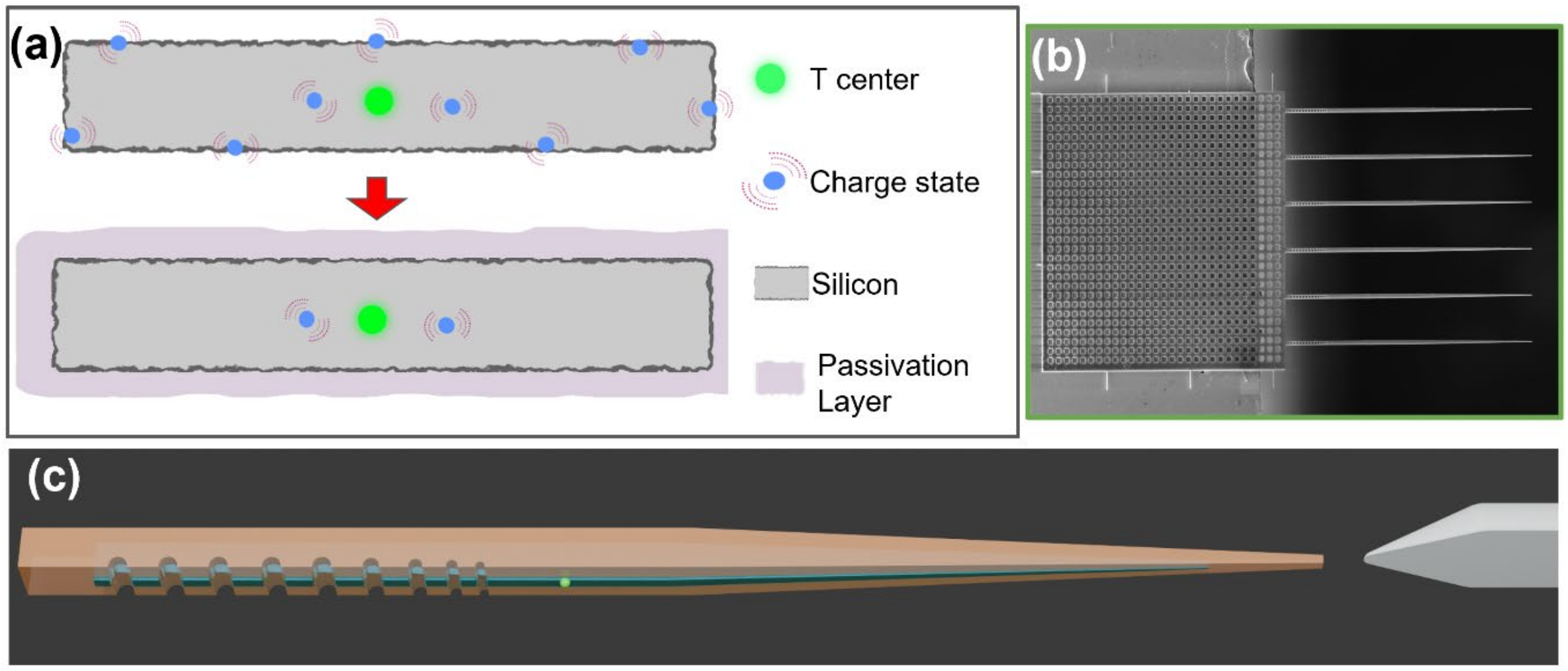

Figure 1: (a) Schematic of a T center in nanophotonic devices before and after surface passivation. (b) Scanning electron micrograph of the transfer printed structure, which has a pad with an array of six nanobeams. (c) Schematic of a cross section of a nanobeam mirror waveguide coupled to a lensed fiber, the waveguide is coated with $Al_2O_3$.

Figure 1(a) illustrates surface passivation of a T center in nanophotonic devices. Charge noise at the silicon surface and at trap sites within the crystal can produce slowly varying electric fields at the emitter, resulting in spectral diffusion of the optical transition. Because T centers in nanophotonic structures are usually 100–200 nm away from the silicon surfaces, surface-related charge fluctuations constitute a major source of spectral broadening. Here, we passivate the exposed silicon surfaces with a thin $Al_2O_3$ layer, thereby reducing the influence of surface charge states on the local environment of the T centers.

To efficiently collect emission from individual T centers, we use tapered silicon nanobeam waveguides designed to couple the emitter emission into a guided optical mode and subsequently into a lensed fiber [9]. We created T centers in silicon-on-insulator substrates by ion implantation followed by thermal annealing and fabricated the nanobeam structures using electron-beam lithography, dry etching, and wet etching. The nanobeams were subsequently transfer-printed onto a separate carrier chip to form cantilevered structures that enable direct coupling to a lensed fiber. A representative scanning electron micrograph of the fabricated nanobeam waveguide is shown in Figure 1(b). Further details of the T center formation and nanobeam fabrication processes are provided in our previous report (Ref. [9] ).

We deposited $Al_2O_3$ conformally onto the suspended nanobeam structures using a Beneq atomic layer deposition (ALD) system, as illustrated in Figure 1(c). The deposition was performed at 150 °C using trimethylaluminum and water as the aluminum and oxygen precursors, respectively. A silicon witness sample was co-loaded with the nanophotonic devices during each deposition, and the resulting $Al_2O_3$ thickness was subsequently determined by spectroscopic ellipsometry.

We carried out the optical measurements in a fiber-coupled cryogenic probe station at 1.74 K. The fiber probe station enables high-efficiency coupling between lensed fiber and the nanobeam, with a collection efficiency of 70%. In order to characterize the optical linewidth, we performed photoluminescence excitation (PLE) measurements by resonantly exciting the T centers at zero phonon line with a tunable laser in the telecom O-band while detecting the corresponding phonon sideband emission using a superconducting nanowire single photon detector (SNSPD) and time-correlated single photon counter (TCSPC). We measured time-resolved photoluminescence to investigate excited state lifetimes of the T centers with different passivation conditions. In order to perform additional above-bandgap illumination measurement, we employed an above-bandgap laser at a wavelength of 980 nm. We utilized two tunable O-band lasers for a spectral hole burning experiment. Details of the setup are provided in the Supplementary Information section 1.

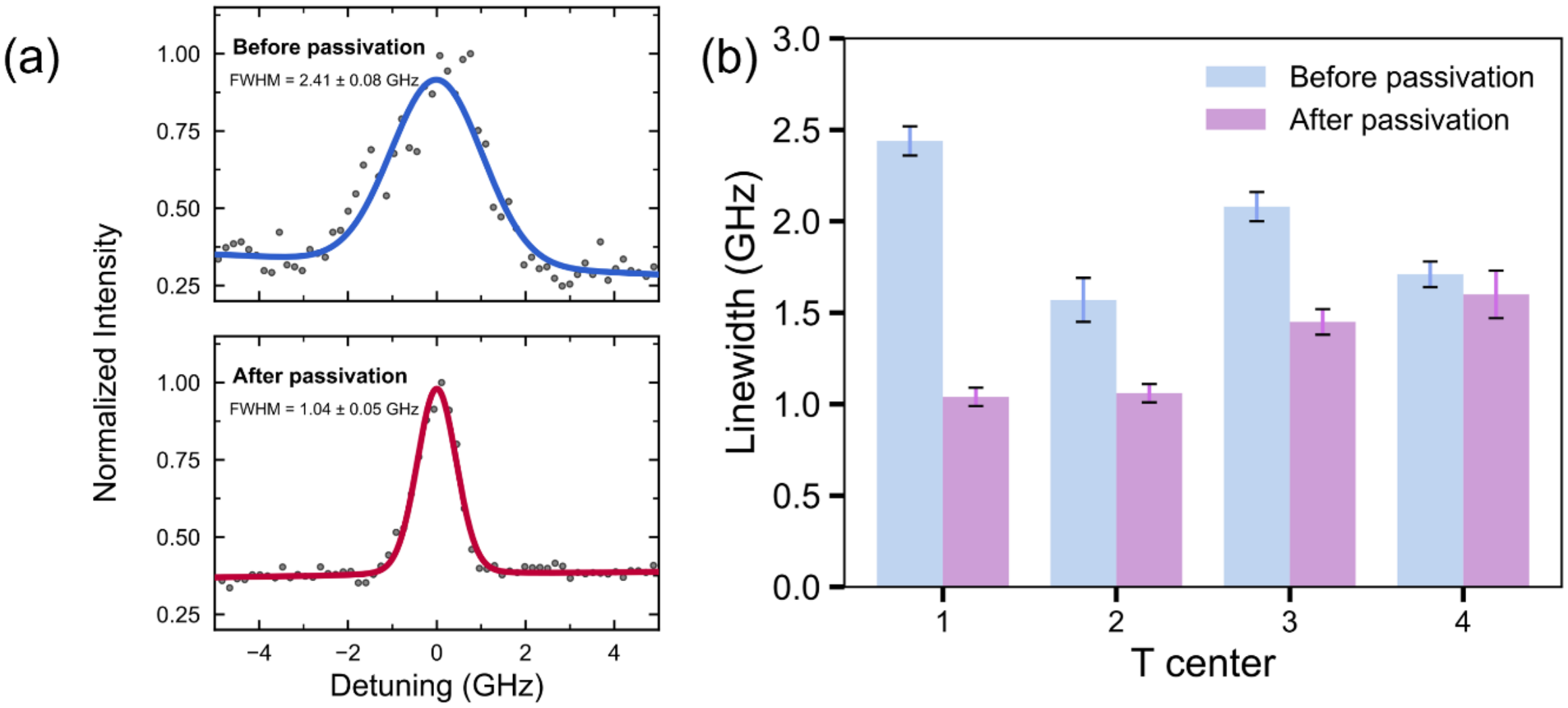

Figure 2. (a) Representative PLE spectra of a T center before (blue) and after (red) $Al_2O_3$ passivation. Solid lines are Voigt fits to the data. (b) Linewidths extracted from the PLE measurements of four different T centers before and after passivation.

For evaluating the effectiveness of the $Al_2O_3$ surface passivation, we compared the optical linewidth of the T center zero-phonon line before and after passivation. After the initial measurements, we removed the sample from the cryostat, deposited 11 nm of $Al_2O_3$ by ALD, and repeated the measurements on the same emitters by returning to their original locations on the nanobeam. Figure 2(a) shows PLE spectra of the same T center before and after $Al_2O_3$ surface passivation. All spectra were acquired at a pump power of $0.1P_{sat}$, where $P_{sat}$ is the saturation pump power extracted from the pump-power dependence. Fitting the PLE spectra to Voigt profiles yielded the optical linewidths. For this emitter, the linewidth decreased from 2.41 GHz to 1.04 GHz after passivation, corresponding to a 57% reduction. We performed the same before-and-after comparison on three additional emitters, as summarized in Figure 2(b). All four emitters showed linewidth narrowing after passivation, with an average reduction of 32%. This level of improvement is comparable to linewidth reductions reported for surface-passivated quantum dots, where surface treatment has similarly been used to suppress charge-noise-induced spectral broadening [18]. The emitter-to-emitter variation in linewidth reduction may reflect differences in proximity to the surface and in the relative weight of surface charge noise versus charge fluctuations in the bulk silicon. Emitters dominated by surface charge states are therefore expected to benefit most from passivation.

Next, we varied the $Al_2O_3$ thickness systematically and measured the resulting T center optical properties to identify the optimal value. We studied $Al_2O_3$ films of 7, 11, 15, 20, and 26 nm, measuring the optical linewidths of ten emitters at each thickness. Second-order autocorrelation measurements confirm that these T centers are single emitters (Supplementary Information section 2). The statistical analysis, shown in Figure 3(a), indicates that the linewidth reduction saturates near 11 nm, suggesting that additional $Al_2O_3$ thickness does not further suppress spectral diffusion from surface charge states. $Al_2O_3$ passivates the silicon surface by terminating dangling bonds and by providing a fixed charge layer at the interface, which stabilizes the fluctuating surface charges. A minimum thickness is required for continuous coverage; below it the surface is only partially passivated, while above it additional material does not further suppress the surface charge fluctuations.

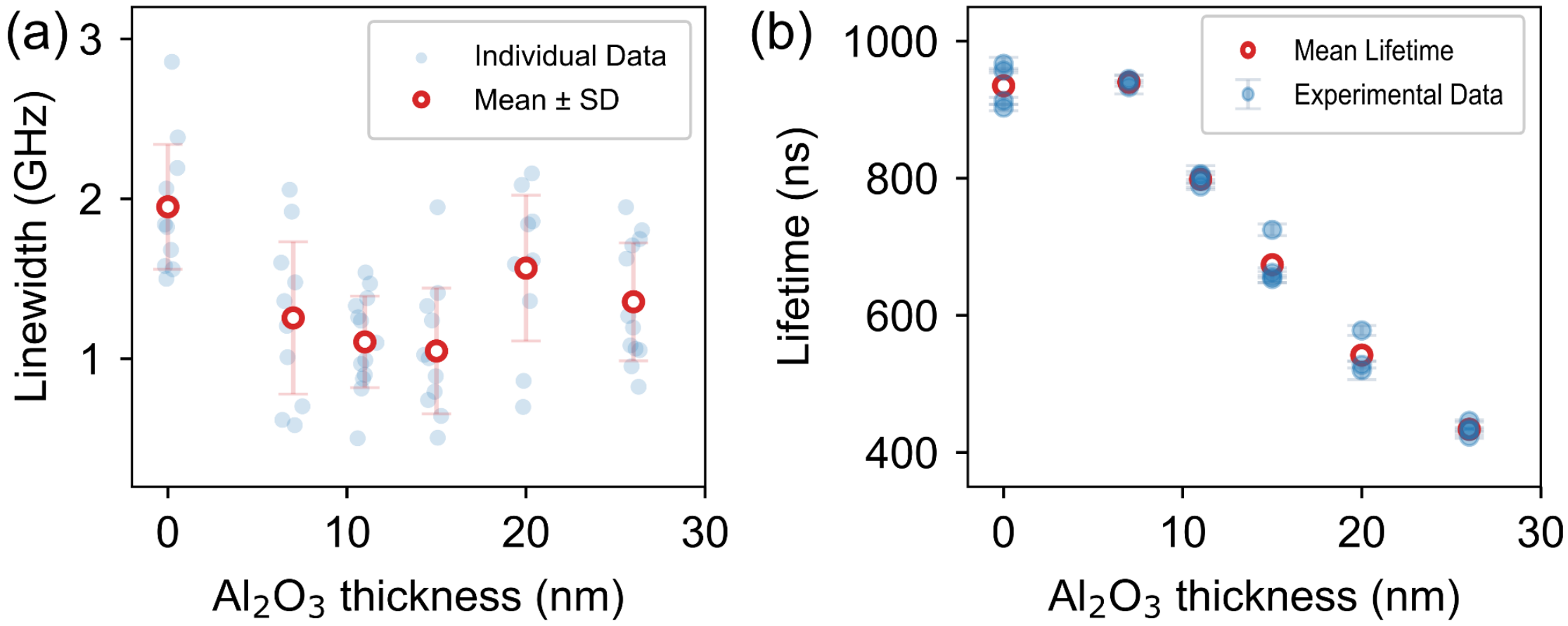

Figure 3. Dependence of T center optical properties on $Al_2O_3$ passivation thickness. (a) Mean optical linewidth of T centers as a function of $Al_2O_3$ thickness. Red circles represent the average linewidth obtained from Voigt fits to individual photoluminescence excitation spectra, with error

bars denoting the standard deviation. Light blue circles represent the linewidths of individual emitters at each passivation thickness. (b) Excited-state lifetime as a function of $Al_2O_3$ thickness. Red circles represent the mean excited-state lifetime at each passivation thickness, while blue circles correspond to lifetimes extracted from exponential decay fits to individual emitters. Error bars represent the uncertainty in the fitted lifetime obtained from monoexponential fitting of the decay curves.

To determine whether surface passivation introduces additional non-radiative decay channels, we measured the excited-state lifetime of four emitters at each $Al_2O_3$ thickness. The lifetime decreases with increasing $Al_2O_3$ thickness, as shown in Figure 3(b). We attribute the lifetime reduction at larger $Al_2O_3$ thicknesses to strain-related enhancement of non-radiative decay, likely through activation or modification of local defect and trap states, although direct structural measurements are required to identify the microscopic mechanism [23–26]. This assignment is supported by a concurrent redshift of the T center emission wavelength with increasing $Al_2O_3$ thickness (Supplementary Information section 3), indicating that thicker $Al_2O_3$ films impose greater strain on the silicon nanostructure. Taken together, these results identify 11 nm as the optimal $Al_2O_3$ thickness in our devices that provides near-saturated linewidth improvement while minimizing strain-induced non-radiative decay.

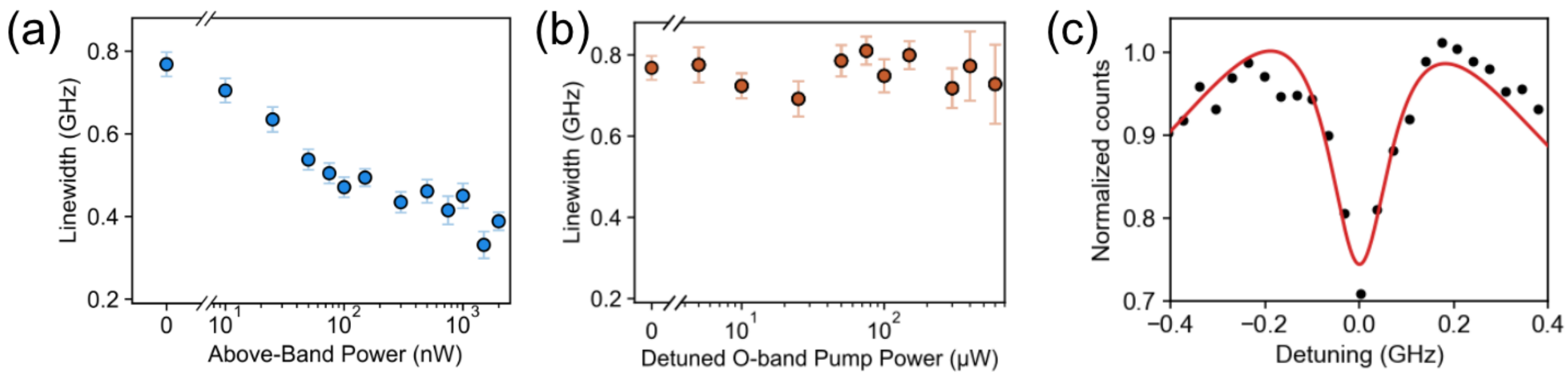

Figure 4. (a) Effect of above-bandgap excitation (980 nm laser) on optical linewidth, the linewidth is obtained from a Voigt fitting of PLE data. (b) Optical linewidth of a T center with different powers of a second O-band laser blue-detuned by 0.6 nm from the T center emission wavelength. (c) Spectral-hole burning data with probe power = $0.06P_{sat}$ and pump power = $0.26P_{sat}$.

Although $Al_2O_3$ passivation substantially narrows the T-center optical transition, the linewidth is still far above the transform limit, indicating that additional broadening mechanisms remain. To test whether this residual broadening is sensitive to the local charge environment, we measured the PLE linewidth of a passivated T center while applying continuous-wave 980-nm above-bandgap illumination (Figure 4(a)). As the 980 nm power increased, the linewidth decreased by roughly a factor of two, reaching a minimum of 0.3–0.4 GHz. In contrast, applying an additional detuned O-band laser produced no systematic narrowing over the measured power range (Figure 4(b)). In both Figure 4(a) and 4(b), the probe O-band laser power was held at 0.15 $P_{sat}$. These observations are consistent with photogenerated carriers filling nearby charge traps and partially stabilizing the local electric-field environment. [27,28]

To separate the residual slow spectral diffusion from homogeneous broadening, which acts on timescales comparable to the optical lifetime, we performed spectral-hole-burning measurements on a T-center nanobeam passivated with a 15-nm-thick $Al_2O_3$ layer. The measurement used two tunable O-band lasers serving as pump and probe, respectively. The pump laser was fixed near the center of the zero-phonon line, while the probe laser was scanned across the transition. When the two lasers addressed the same instantaneous optical resonance, saturation by the pump suppressed the probe-induced fluorescence and produced a spectral hole. Fitting the spectrum with a Lorentzian hole superimposed on a broad Gaussian PLE background gives a hole full width at half maximum (FWHM) of $150.8 \pm 22.0$ MHz. Because the spectral-hole FWHM in the low-power limit is twice the homogeneous linewidth, and finite optical power broadens the hole further, this measurement places an upper bound of approximately 75 MHz on the homogeneous linewidth. The substantially larger PLE linewidth therefore indicates that slow spectral diffusion remains the primary contribution to the residual broadening.

Our results show that $Al_2O_3$ surface passivation suppresses a substantial component of the spectral diffusion of T centers in nanophotonic devices, but does not eliminate the residual charge-induced broadening. Increasing the thickness beyond approximately 11 nm provides little additional linewidth narrowing while reducing the excited-state lifetime, indicating that further improvement will require optimization of the passivation interface rather than simply increasing the coating thickness. We could employ a surface treatment using HF:HCl solution to preterminate the silicon dangling bonds[20,29], and post deposition annealing for optimum activation of the surface passivation layer[30]. The additional narrowing observed under above-bandgap illumination suggests that charge traps not fully stabilized by the $Al_2O_3$ treatment also contribute to the remaining spectral diffusion. Complementary approaches, such as reducing fabrication-induced surface damage through atomic layer etching[13,31] or electrically stabilizing the local charge environment using diode structures[32,33], could therefore be integrated with optimized surface passivation to further narrow the optical transition. Combined with cavity integration to accelerate the spontaneous emission rate, these strategies chart a path toward indistinguishable photon generation and scalable spin–photon interfaces based on silicon T centers.

In conclusion, we demonstrate that $Al_2O_3$ surface passivation provides an effective route to reduce optical linewidth broadening in nanophotonically integrated silicon T centers. By comparing individual emitters before and after passivation, we recorded an average linewidth narrowing of 32%, with actual reductions reaching up to 57%. Systematic thickness variations reveal an optimal passivation layer of 11 nm, beyond which strain-induced non-radiative decay begins to negatively impact the excited-state lifetime without offering further linewidth reduction. Above-bandgap illumination further narrows the PLE linewidth to approximately 400 MHz, while spectral hole burning bounds the homogeneous linewidth at approximately 75 MHz, showing that the slow spectral diffusion remains after passivation. Our results establish $Al_2O_3$ surface passivation as a practical, CMOS-compatible strategy for mitigating spectral diffusion in silicon T centers.

## Author Information

### Corresponding authors

*Chang-Min Lee, cmlee@umd.edu
*Edo Waks, edowaks@umd.edu

### Notes

While preparing this manuscript, we became aware of the related work by Johnston *et al.*[34] .

## Supporting Information

Details of experimental setup, additional data on second-order autocorrelation measurement, and effect of $Al_2O_3$ on T center emission wavelength.

## Acknowledgements

The authors acknowledge Jasvith Raj Basani, Carolina Crosta, Yuxi Jiang, and Abhijit Biswas for technical assistance and helpful discussions. The authors thank Dr. Nam Kim of the University of Maryland NanoCenter and Dr. Joshua Thompson of the Laboratory for Physical Sciences for their technical support in sample preparation. The authors would like to acknowledge financial support from the National Science Foundation (grant #ECCS2423788), the Department of Energy (grant #DESC0026071), and the Air Force Office of Scientific Research (grant #FA95502310667 and #FA95502410266).

# Supplementary Information for

# Surface passivation for narrowing optical linewidth of silicon T centers in nanophotonic devices

Fariba Islam[1,2,3], Chang-Min Lee[1,2,*], Kyu-Young Kim[1,2,4], Sorah Fischer[1,2,3], Purbita Purkayastha[1,2,5], Amirehsan Alizadehherfati[1,2,3], and Edo Waks[1,2,3,5,*]

[1] Institute for Research in Electronics and Applied Physics, University of Maryland, College Park, Maryland 20742, USA
[2] Joint Quantum Institute, University of Maryland, College Park, Maryland 20742, USA
[3] Department of Electrical and Computer Engineering, University of Maryland, College Park, Maryland 20740, USA
[4] Department of Physics, Ulsan National Institute of Science and Technology, Ulsan 44919, Republic of Korea
[5] Department of Physics, University of Maryland, College Park, Maryland 20742, USA

## Section 1. Measurement setup

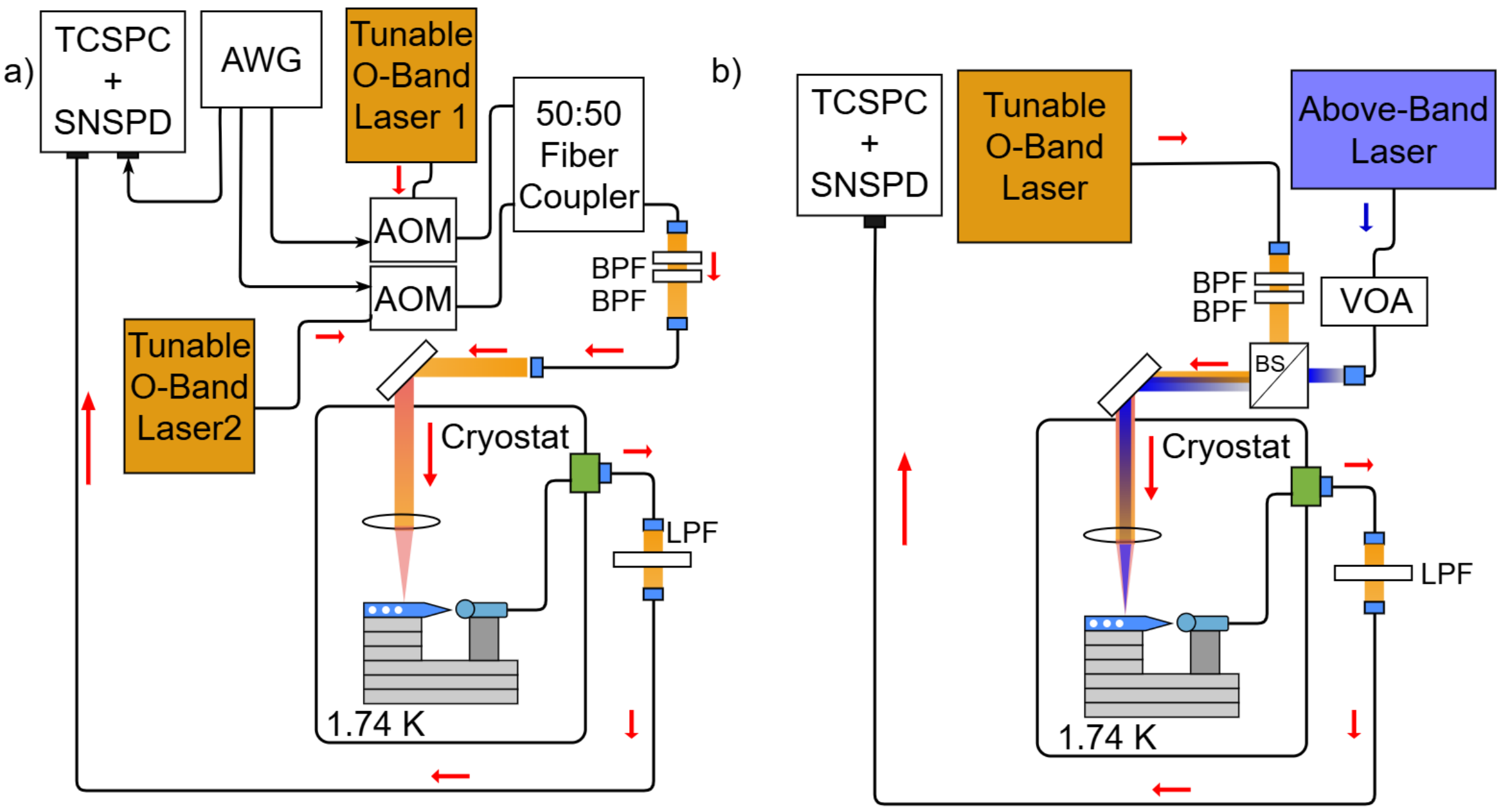

Figure S1. (a) Measurement setup for photoluminescence excitation (PLE), time-resolved lifetime measurement, and spectral hole burning. TCSPC: time-correlated single photon counter, SNSPD: superconducting nanowire single photon detector, AWG: arbitrary waveform generator, AOM: acousto-optic modulator, BPF: band-pass filter, LPF: long-pass filter (b) Set-up for observing the effect of above-band power on optical linewidth. VOA: variable optical attenuator, BS: beam-splitter.

We used the experimental setup shown in Figure S1(a) for photoluminescence-excitation (PLE), time-resolved PLE, Hanbury Brown–Twiss (HBT) measurements, and spectral hole burning. In the first three measurements we kept only one tunable O-band laser on. For time-resolved PLE, the excitation laser was pulsed using an acousto-optic modulator (AOM) controlled by an arbitrary waveform generator (AWG). When the AWG was operated with a constant DC bias, the AOM transmitted continuous-wave excitation. The excitation laser was passed through two band-pass filters with a center wavelength of 1330 nm and a bandwidth of 12 nm to suppress unwanted sideband emission. Photoluminescence from the nanobeam waveguide was collected using a lensed fiber and directed through a long-pass filter with a cutoff wavelength of 1350 nm before reaching the superconducting nanowire single-photon detector (SNSPD). This filtering configuration allowed us to selectively detect the phonon-assisted sideband emission from the T centers.

To perform the spectral hole burning measurement, we used two tunable O-band lasers and combined them using a 50:50 fiber coupler and then transmitted them along the same pumping laser path. We varied the pump power keeping the probe power constant at $0.06P_{sat}$. For the Hanbury Brown–Twiss (HBT) measurements, the excitation laser was tuned to the resonant wavelength of the T center, as determined from the PLE spectrum. The collected emission was divided using a 50:50 fiber coupler and directed to two independent superconducting nanowire single-photon detector (SNSPD) channels. The photon arrival times recorded by the two detectors were then correlated to obtain the second-order intensity autocorrelation function, $g^2(\tau)$.

We utilized the setup in Figure S1(b) to observe the above-band laser effect on the linewidth. A laser diode with a wavelength of 980 nm was employed as an additional pumping laser. The two different wavelength lasers were mixed using a beamsplitter and sent to the sample and we collected the sideband as described earlier.

## Section 2. Confirmation of single emitter via second order autocorrelation measurements

To confirm that the emission peaks observed in the PLE spectra originate from single emitters, we randomly selected four emitters from our sample before atomic layer deposition and performed autocorrelation measurements as described in Section 1. All measurements exhibited a background-corrected value of $g^2(0) < 0.5$, confirming that the emissions originated from individual T centers. A representative autocorrelation measurement histogram is shown in Figure S2. The measurement was performed at 6 K using a pump power of $P = 0.35P_{sat}$. The autocorrelation data were background-corrected and fitted following the analysis procedure described in our previous work (Ref. [1]).

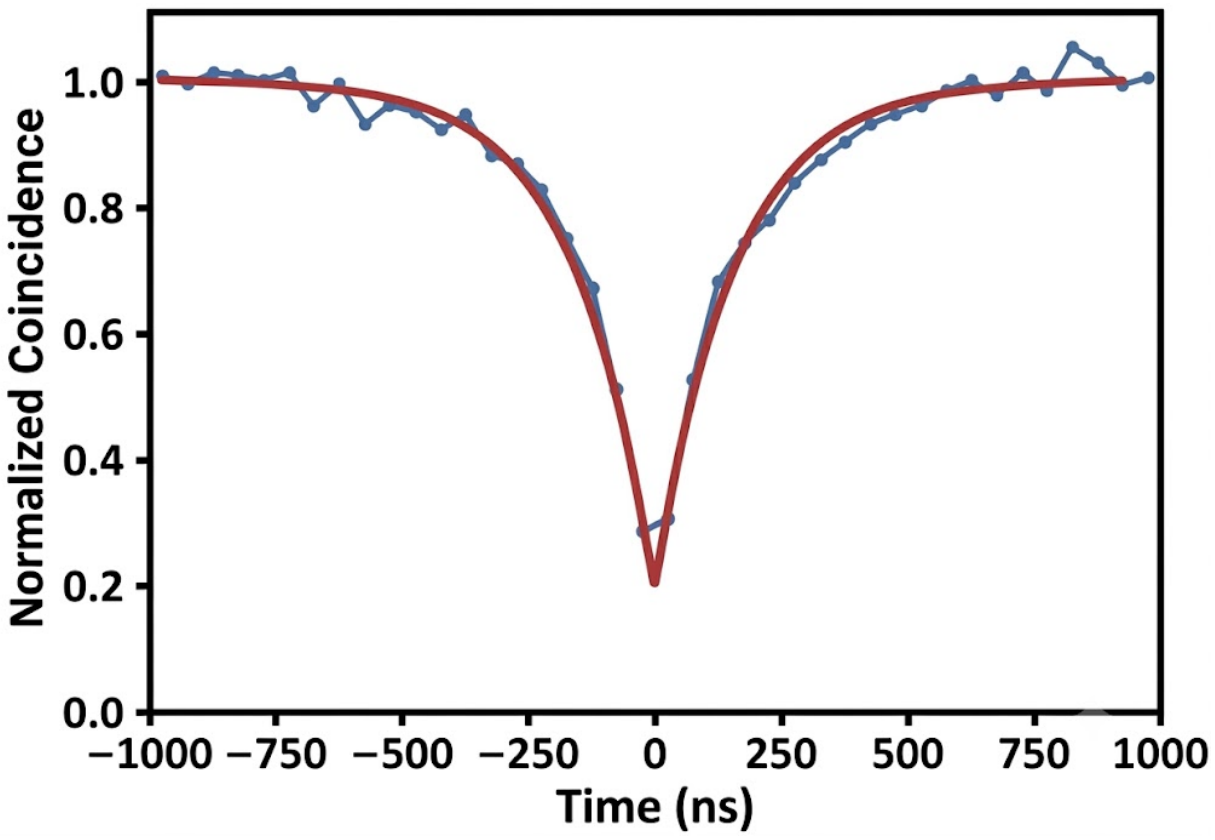

Figure S2. A representative background-corrected $g^2(\tau)$ trace with a fit to a single-exponential antibunching model where $g^2(0)=0.20$.

## Section 3. Effect of increasing the thickness of $Al_2O_3$ film on emission wavelength of T center

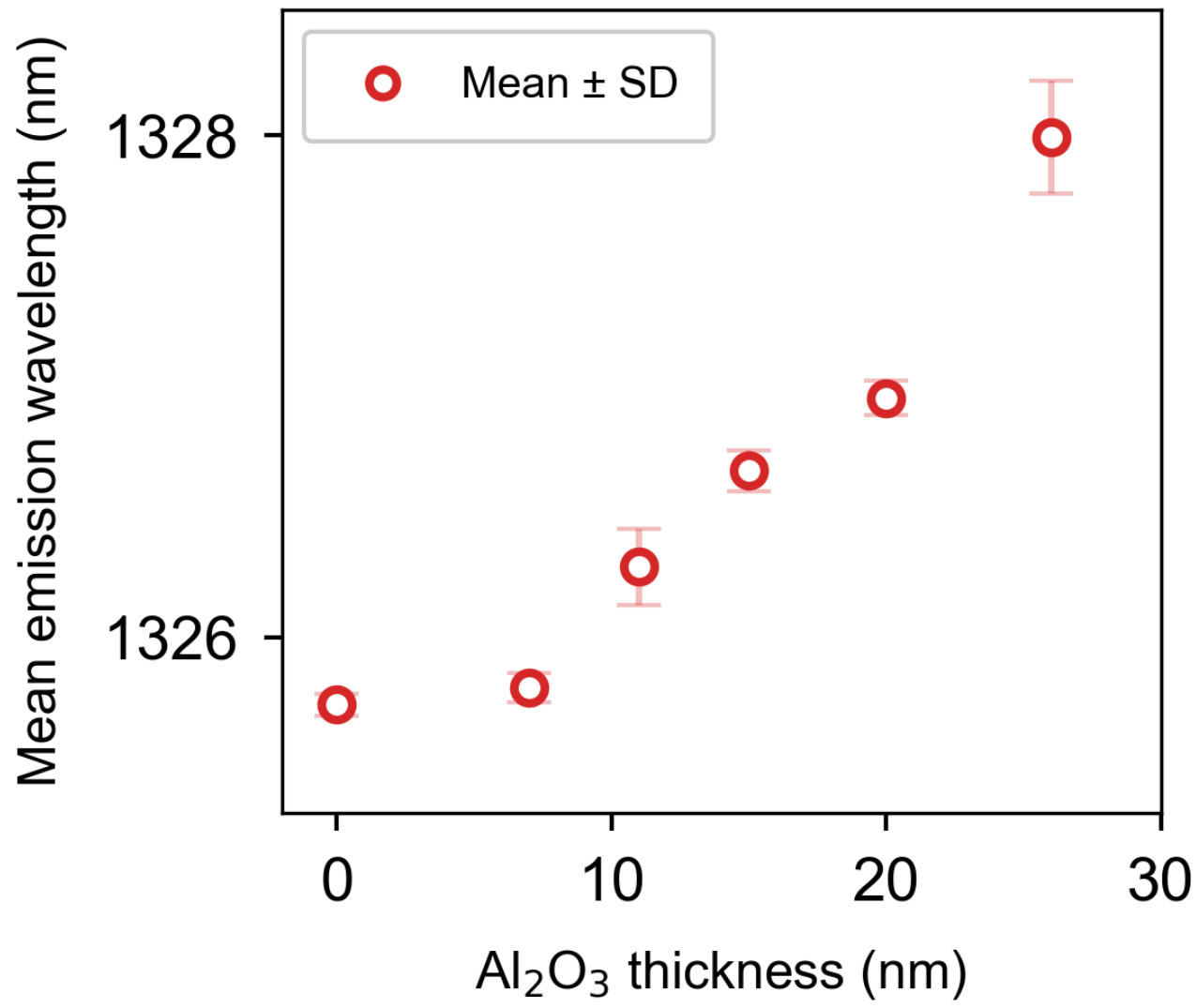

Figure S3. Emission wavelength shift with increase in thickness of the $Al_2O_3$ film. The emission wavelength is obtained through lorentzian fit of the PLE data.

From the statistical analysis of the PLE linewidth at different $Al_2O_3$ thickness data from which we extracted Figure 3 in the main manuscript, we can also extract emission wavelengths as a function of $Al_2O_3$ thickness. As shown in Figure S3, the emission wavelength exhibits an overall redshift with increasing $Al_2O_3$ thickness, indicating a strain-induced shift of the T center optical transition.